# Tuning Cu/Diamond Interfacial Thermal Conductance via Nitrogen-Termination Engineering

Guang Yang[1], Xinling Tang[1,*], Zhongkang Lin[1], Yulin Gu[1], Wei Hao[1], Yujie Du[1], Xiaoguang Wei[1,2,*]

[1] Beijing Huairou Laboratory, Beijing 101407, China

[2] China Southern Power Grid Novel Electric Power System (BEIJING) Research Institute Co., Ltd, Beijing, China

*Author to whom any correspondence should be addressed.

**E-mails:** tangxinling@neps.hrl.ac.cn and weixiaoguang@neps.hrl.ac.cn

# Abstract

Cu-diamond composites are recognized as promising high-thermal-conductivity candidates for electronic cooling, offering tunable properties and competitive cost. However, their performance is significantly limited by the poor Cu/diamond interfacial thermal conductance (ITC). Here, we propose a nitrogen-termination strategy to tune the ITC of Cu/diamond interfaces and unravel atomistic mechanisms by which nitride interlayers tailor phonon transport. Based on the MACE machine-learning interatomic potential (MLIP) framework, we fine-tune the pre-trained `MACE-MPA-0` foundation model by incorporating customized C-N-Cu training datasets. Through MLIP-driven lattice dynamics simulations, we demonstrate that an atomically flat N-termination on diamond enhances the ITC by 21% compared to the bare Cu/diamond interface. Mode-resolved phonon spectroscopy reveals that the LA phonons with frequency above 4 THz and wavevectors near $\Gamma$-$X$ and $\Gamma$-$U$ directions are selectively modulated by N-termination engineering. Analyses of local vibrational states and interfacial bonding further indicate that the N-termination on diamond tunes the interfacial heat conduction via surficial mass modification and bonding regulation, as evidenced by variations in LDOS overlap and COHP spectra. These findings open venues for tuning heat transfer across Cu/diamond interfaces via non-metallic modification—which avoids the graphitization issues associated with metallic coatings—and provide novel guidelines for upgrading the phonon-mediated heat transfer in Cu-diamond composites.

# 1. Introduction

The continuous advancement of power electronics and information technology demands that modern electronic devices handle increasingly high power densities. However, these performance gains intensify heat dissipation challenges, which critically influence device performance and long-term reliability[1–8]. Efficient heat dissipation relies on both the intrinsic thermal conductivity ($\kappa$) of constituent materials and the interfacial thermal conductance (ITC), thus developing high-$\kappa$ packaging materials while simultaneously creating low-thermal-resistance interfaces[1,9–14] is essential for addressing thermal management challenges[13,15–20]. Although diamond and other emerging chemical compounds—such as BAs[1,21–25], $c$-BN[19], $\theta$-TaN[26–28]—exhibit exceptional thermal conductivity, their large-scale application is yet constrained by complex synthesis processes and prohibitively high costs. These difficulties have shifted research focus toward composite material systems as a more promising and technologically viable solution for high-$\kappa$ packaging and thermal management.

High-$\kappa$ composites are categorized into metal-matrix composites (metallic matrices and non-metallic reinforcements) and non-metallic-matrix composites. Among them, metal-diamond composites, particularly the Cu-diamond systems, have been widely adopted due to some distinct advantages[29]: (i) high effective thermal conductivity, with $\kappa$ values reaching 600 W/(m·K); (ii) tunable coefficients of thermal expansion, which can be matched to those of semiconductors (4–7×10$^{-6}$ K$^{-1}$); and (iii) simpler processing and more controllable cost compared to bulk diamond. Cu-diamond composites are usually fabricated by the pressure infiltration process or the spark plasma sintering (SPS) technique. Nevertheless, the inherent chemical inertness, poor interfacial wettability and weak phonon spectra matching between the Cu matrix with diamond particles collectively severely impede interfacial thermal transport[29–31], resulting in ITC values even below 20 MW/(m$^2$·K). This low ITC leads to a substantial reduction in the overall thermal performance of Cu-diamond composites[32–36].

To tackle this challenge, various interface modification strategies have been proposed, primarily involving matrix alloying or surface modification of the diamond reinforcements. Diamond surface modification, in particular, has been extensively investigated as an effective approach to upgrade the thermal properties of Cu-diamond composites[29,30,34,36–40]. In practice, metallic coatings (e.g., Ti, Cr, Zr, W, Mo) or non-metallic coatings (e.g., B, Si, SiC) are deposited onto diamond particles at the micro- or nanoscale via techniques such as chemical vapor deposition (CVD), physical vapor deposition (PVD), or sputtering, etc. These modification coatings help enhance ITC by providing phonon bridging and simultaneously improve the wettability between molten Cu and diamond. For example, various carbide interlayers formed during processing (e.g., TiC, ZrC, WC, Cr$_3$C$_2$, SiC and B$_4$C) that generally reduce the wetting angles with liquid Cu while improving phonon transmission efficiency. Notably, some recent studies reported that the thermal conductivity of Cu-diamond composite could exceed 700 W/(m·K) with the aid of Zr-ZrC or W-(Zr, W)C interlayers[29,30].

Although metallic modifications are commonly employed in the fabrication of Cu-diamond composites, it is well documented that transition metals (e.g., Co, Cr, Ni, Mn, Fe, and Cu) can catalyze the spontaneous conversion of diamond ($sp^3$) to graphite phase ($sp^2$) at elevated temperatures (>700 °C)[41–43]. This graphitization effect significantly degrades the interfacial heat conduction. For instance, Wang et al.[41] introduced a TiC/$Cr_3C_2$ gradient interlayer at the Cu/diamond interface to enhance phonon transmission, yet residual graphite formation and excessive thickness of $Cr_3C_2$ interlayer together reduced the composite's thermal conductivity to only 549 W/(m·K). In contrast, non-metallic coatings do not induce such diamond-to-graphite transformation, offering a promising route to circumvent this issue fundamentally.

Nitrogen (N) is the most ubiquitous foreign element in diamond, owing to its similar atomic radius and valence-shell configuration, which facilitates substitution for carbon atoms[44]. These nitride impurities function as both disruptive contaminants and functional assets, dictating the diamond's color, thermal conductivity, electrical performance, and quantum coherence[45]. In the functionalization of diamond, nitrogen termination (N-termination), an emerging topic in diamond electronics and quantum photonics, has been considered as an effective surface-bonding state. Compared to the hydrogen or oxygen terminations, N-termination engineering offers superior capability for tuning shallow color centers, improving optical stability, enhancing quantum coherence, or constructing antiscaling coatings[44,46–49]. The developing N-termination engineering inspires our investigation into its potential for tuning the ITC of Cu/diamond interface without deteriorating the diamond's $sp^3$ structure. This approach is supported by the strong chemical affinities of both C–N and N–Cu bonds[50,51] and the technical feasibility of N-termination processes. However, compared with the foregoing metallic coatings, studies on non-metallic modifications, especially nitrogen-based surface engineering for Cu-diamond composites remains absent. Its potential for tuning the Cu/diamond ITC by modulating the bonding strength and intermediate mass has not been exploited.

Here, we focused on the nitrogen termination engineering of diamond surfaces to explore its potential for tuning the Cu/diamond interfacial thermal conductance. First, within the MACE framework, we finetuned the pre-trained `MACE-MPA-0` foundation potential using customized C-N-Cu configurations to obtain a quantum-accurate potential, namely the `MACE-MPA-ft` model. We performed lattice dynamics simulations based on `MACE-MPA-ft` model to investigate how the nitrogen composition on the diamond surface influences interfacial thermal transport. Then, we conducted analyses of local vibrational states and interfacial bonding to unveil the atomistic mechanisms by which nitrogen termination tunes the Cu/diamond interfacial thermal conductance, which will be helpful to guide the modulation of phonon-mediated heat transfer in Cu-diamond composites.

## 2. Methods

### 2.1. The MACE MLIP framework

Machine-learned interatomic potentials (MLIPs) provide representations of the given quantum-mechanical potential-energy surfaces. The MLIP models are generally based on a local and atom-wise decomposition of the total energy[25,52–54],

$$E = \sum_i \varepsilon(\boldsymbol{\sigma}_i), \tag{1}$$

where the atomic energies $\varepsilon$ are learned as a function of the atom's local environment, described by a general structural descriptor $\boldsymbol{\sigma}_i$. The ML-predicted total energy $E$, is then obtained by summing over the per-atom contributions. For message-passing neural network (MPNN) interatomic potential models, structural descriptor $\boldsymbol{\sigma}_i$ can be represented by a node state in the $t$-th layer, i.e., a tuple $\boldsymbol{\sigma}_i^{(t)} = \left(\mathbf{r}_i, z_i, h_i^{(t)}\right)$. Then the representation of atomic energy $\varepsilon$ is,

$$\begin{cases} \varepsilon(\boldsymbol{\sigma}_i) = \sum_{t=1}^{T} R_t\left(\mathbf{r}_i, z_i, h_i^{(t)}\right), \\ h_i^{(t+1)} = U_t\left(\boldsymbol{\sigma}_i^{(t)}, m_i^{(t)}\right), \\ m_i^{(t)} = \bigoplus_{j \in \mathcal{N}(i)} M_t\left(\boldsymbol{\sigma}_i^{(t)}, \boldsymbol{\sigma}_j^{(t)}\right), \end{cases} \tag{2}$$

where learnable readout functions $R_t$ map the node states $\boldsymbol{\sigma}_i^{(t)}$ to atomic energy $\varepsilon$, and $U_t$ is a learnable update function to update the message $m_i^{(t)}$ into new features $h_i^{(t+1)}$. $M_t$ is a learnable message function and $\bigoplus_{j \in \mathcal{N}(i)}$ is a permutation invariant pooling operation over neighbors of atom $i$.

The MACE architecture[54–57], a refined MPNN-based model, is grounded in equivariant message-passing graph tensor networks that retain key geometrical and physical symmetries of atomic structures. The central improvement of MACE is upgrading the $\bigoplus_{j \in \mathcal{N}(i)} M_t$ in **Eq. (2)** to enclose higher body-order equivariant features, referring to the atomic cluster expansion (ACE) paradigm[58,59],

$$\begin{aligned} m_i^{(t)} = & \sum_j u_1\left(\boldsymbol{\sigma}_i^{(t)}; \boldsymbol{\sigma}_j^{(t)}\right) + \sum_{j_1, j_2} u_2\left(\boldsymbol{\sigma}_i^{(t)}; \boldsymbol{\sigma}_{j_1}^{(t)}, \boldsymbol{\sigma}_{j_2}^{(t)}\right) + \\ & \cdots + \sum_{j_1, \dots, j_\mu} u_\mu\left(\boldsymbol{\sigma}_i^{(t)}; \boldsymbol{\sigma}_{j_1}^{(t)}, \dots, \boldsymbol{\sigma}_{j_\mu}^{(t)}\right), \end{aligned} \tag{3}$$

in which the $u$ functions are learnable, the sums run over the neighbors of $i$, and $\mu$ is a hyper-parameter corresponding to the maximum correlation order (body order minus 1) of the message function $M_t$ with respect to the node states $\boldsymbol{\sigma}_i^{(t)}$. The MACE framework has transformed the atomistic modelling of materials by enabling simulations of *ab initio* quality with accelerated implementation[56,60].

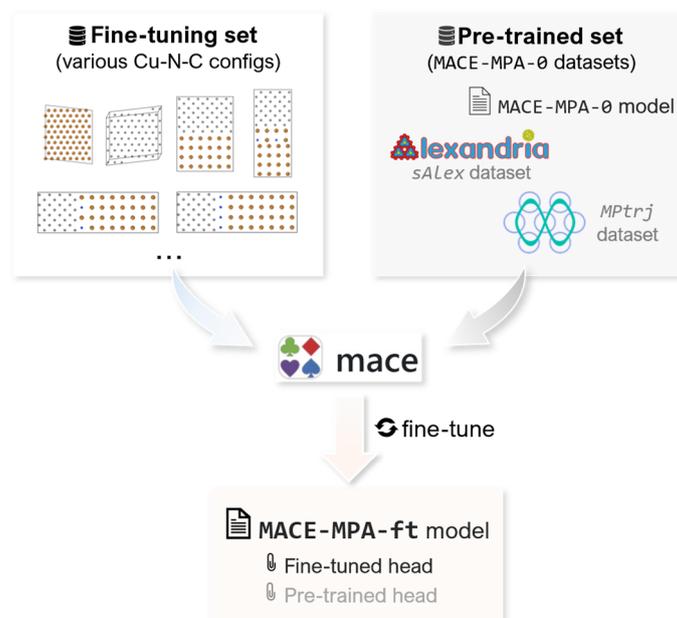

**Fig. 1.** Overview of model finetuning (ft) processes in MACE based on the pre-trained `MACE-MPA-0` foundation model. The "finetuned" head of the `MACE-MPA-ft` model is used for modelling and phonon analysis in this study.

However, application-specific MLIPs are often constrained by two major limitations: (i) the substantial computational and human effort required to develop and validate potentials for each particular system, and (ii) the poor transferability across different chemical environments. Recent advances have demonstrated that general-purpose atomistic MLIP models, trained on publicly available datasets of large size, can perform stable molecular dynamics simulations across a broad spectrum of molecules and materials[60]. Within the MACE framework, a series of such general-purpose models, represented by the pre-trained `MACE-MP-0` and `MACE-MPA-0` models[60], can be readily applied as starting or "foundation" potentials for any atomistic system of interest. These foundation models can be conveniently finetuned with only a small set of application-specific data to reach *ab initio* accuracy when desired. The development of robust, broadly applicable force-field models fundamentally changes atomistic modelling: experienced researchers can obtain reliable results more rapidly, while the entry barrier for newcomers is significantly lowered.

While the broad applicability of the `MACE-MPA-0` foundation model demonstrates its robustness, its out-of-the-box accuracy remains insufficient to rival or replace *ab initio* calculations for some specific cases such as lattice dynamics and phonon transport simulations[60]. To address this limitation, we fine-tuned the

`MACE-MPA-0` model based on customized C-N-Cu datasets. We collected ~300 Cu/nitride-interlayer/diamond heterostructure configurations as the fine-tuning inputs (detailed in Section 3.1). The fine-tuning process incorporates a subset of the original training data of `MACE-MPA-0` into the loss function while optimizing on the new data. This procedure yielded the tailored `MACE-MPA-ft` potential, with the overall workflow illustrated in **Fig. 1**. Previous studies have validated that such finetuned models can substantially reduce errors in energy, forces, and stresses on the target datasets compared to both the original foundation model and even directly trained specific-purpose potentials[60].

## 2.2. Lattice dynamics simulations

Lattice dynamics is applied to study the phonon transport picture and analyze the interfacial thermal conductance in this work. Though lattice dynamics method is based on the harmonic approximation, it is recently reported that the elastic phonon scatterings contribute ~65% to Cu/diamond ITCs regardless of the crystal orientations[38]. Therefore, we focus on modelling the harmonic phonon transmission across interfaces in this work, and the effects of inelastic phonon scatterings will be investigated in our future studies. Similar to atomistic Green's function (AGF) method[61–63], the robust lattice dynamics (RLD) method[64] proposed by Yang et al. also partitions a heterostructure into three parts: the left and right "Lead" regions, and the central "Device". Assuming the amplitude of each incident phonon mode is $A_I = 1$, the modal phonon transmittance $\mathcal{T}_{q\nu}$ is defined as the energy flux ratio of transmission mode $T_{q,\nu}^{(k)}$ to that of incident mode $I_{q,\nu}$. Similarly, the modal reflectance $\mathcal{R}_{q\nu}$ is defined as the ratio of reflected energy flux to incident one,

$$\begin{cases} \mathcal{T}_{q\nu} = \sum_k \frac{v_{T^{(k)},z}}{v_{I,z}} |A_{T^{(k)}}|^2, \\ \mathcal{R}_{q\nu} = \sum_j \frac{-v_{R^{(j)},z}}{v_{I,z}} |A_{R^{(j)}}|^2, \end{cases} \quad (4)$$

where $v_{I,z}$, $v_{T^{(k)},z}$ and $v_{R^{(j)},z}$ denote the group velocities of the incident, transmitted, and reflected phonon modes along the transport ($z$) direction, respectively. $A_{T^{(k)}}$ and $A_{R^{(j)}}$ represent the mode amplitudes of the transmitted and reflected phonons.

The energy conservation law requires that $\mathcal{T}_{q\nu} + \mathcal{R}_{q\nu} = 1$. After deriving the second-order IFCs via finite displacement method, RLD algorithm solves the amplitudes of each bulk phonon mode and directly determines the mode-resolved phonon transmittance and reflectance by Eq. (4). After deriving $\mathcal{T}_{q\nu}$ and $\mathcal{R}_{q\nu}$, the ITCs can be directly calculated according to the Landauer formula[9,61,64,65],

$$h = \frac{1}{V}\sum_{q,v} \hbar\omega_{qv}v_{qv,z}\mathcal{T}_{qv}\frac{\partial f}{\partial T}, \tag{5}$$

where $V$ is the volume of primitive cell, $q$ the phonon wave vector, $v$ the phonon band index, $\omega_{qv}$ the angular frequency of phonon mode $qv$, $v_{qv,z}$ the z-component of phonon group velocity, $f$ the equilibrium Bose-Einstein distribution, and $T$ the temperature. The spectral phonon transmittance can be obtained by a weighted average[64],

$$\mathcal{T}(\omega) = \frac{\sum_{q,v} \hbar\omega_{qv}v_{qv,z}\mathcal{T}_{qv}\frac{\partial f}{\partial T}\delta(\omega_{qv}-\omega)}{\sum_{q,v} \hbar\omega_{qv}v_{qv,z}\frac{\partial f}{\partial T}\delta(\omega_{qv}-\omega)}. \tag{6}$$

Local density of states (LDOS) captures the localized phonon modes near the interface, which is also critical for analyzing the interfacial phonon transmission[9,66]. To calculate LDOS via the RLD method, uniform samplings are performed on phonon modes of the left and right "Lead" regions (in their first Brillouin zones) and regard them as incident phonons. Then, the resulting eigenvectors of atoms $i$ in the "Device" region are computed and the squares of magnitudes of these eigenvectors are summed up over all sampling points, multiplied by a factor of the delta function[64]. Similar to the atom-projected phonon DOS defined within `Phonopy`[67,68] and `ALAMODE`[69], the LDOS of atom $i$ in the device is

$$LDOS_i(\omega) = \sum_{\substack{q \text{ in} \\ \text{left Lead}}} |u_i|^2\delta(\omega-\omega_q) + \sum_{\substack{q' \text{ in} \\ \text{right Lead}}} |u_i|^2\delta(\omega-\omega_{q'}). \tag{7}$$

where the $u_i$ is the displacement of atom $i$ in "Device" region. Compared with classical lattice dynamics, the RLD algorithm incorporates linear algebra transformations and projection gradient descent iterations to rigorously enforce energy conservation, thereby avoiding unphysical phonon transmittance. In contrast to non-equilibrium molecular dynamics (NEMD)[2], RLD offers higher computational efficiency and eliminates non-physical size effects. These advantages make RLD suitable for phonon transport analysis leveraging MLIPs.

### 2.3. Characterization of the interfacial bonding characteristics

It is revealed that bonding strength plays a key role in achieving high thermal conductivity[12,70,71]: materials with strong bonding, such as diamond or other carbon allotropes and *c*-BN tend to have high thermal conductivity[19,72]. Hence, some bonding descriptors, represented by the crystal orbital Hamilton population (COHP) and its integral (ICOHP) are correlate strongly with lattice thermal conduction[70,73]. COHP is defined as partitioning the electronic band structure in terms of the orbital-pair contribution by their Hamiltonian,

$$H_{\mu\nu} = \langle \phi_\mu | \hat{H} | \phi_\nu \rangle, \tag{8}$$

in which $H$ is the Hamiltonian, $\phi_\mu$ is orbital $\mu$, and $\phi_\nu$ is orbital $\nu$. Wavevector **k**-dependent LCAO basis at band $j$ has the following form $\phi_j(\mathbf{k}, r) = c_{j\mu}(\mathbf{k})\phi_\mu(r) + c_{j\nu}(\mathbf{k})\phi_\nu(r) ...$, where $c_{j\mu}$ and $c_{j\nu}$ are the coefficients for orbitals $\mu$ and $\nu$. The coefficients are used to construct the projected density matrix $P_{\mu\nu} = \sum_j^{MOs} f_j c_{j,\mu} c_{j,\nu}$. Energy-dependent COHP can be defined as

$$COHP_{\mu\nu}(E) = H_{\mu\nu} \sum_{j,k} Re(c^*_{\mu,jk} c_{\nu,jk}) \cdot \delta(\varepsilon_j(k) - E). \tag{9}$$

Integrated COHP (ICOHP) is obtained by integration up to the Fermi energy, which can be utilized as a convenient value to characterize chemical bonding.

$$ICOHP_{\mu\nu} = \int_{-\infty}^{\varepsilon_F} COHP_{\mu\nu}(E) dE. \tag{10}$$

Note that the negative value of COHP and ICOHP is used for analysis, since the bonding states in COHP under the Fermi energy are negative. Therefore, the negative value of COHP and ICOHP is the chemical bonding strength descriptor for the material.

Calculation of the COHP and ICOHP is performed using the Local Orbital Basis Suite Towards Electronic Structure Reconstruction (LOBSTER v5.1.1) package[73,74]. These calculations are performed after the self-consistent field (SCF) calculations within VASP[75] that provided the all-electron wavefunction in a plane-wave basis to convert it into LCAO basis. The LCAO basis was generated by projecting chemically intuitive orbitals for each species onto an all electron wave function with a plane-wave basis from VASP. The basis functions and input files were generated with the help of pymatgen[76] package.

## 3. Results and discussion

### 3.1. The finetuned MLIP model

To investigate thermal conduction properties of the Cu/diamond interfaces involving nitrogen, we finetuned the `MACE-MPA-0` foundation model to the customized `MACE-MPA-ft` model with cut-off radius remaining 6.0 Å. Selection of atomic structures in train datasets is critical for constructing reliable MLIP models. As shown in **Fig. 1**, five initial structures, including fcc-Cu, diamond, bare Cu/diamond interface, and the Cu/Cu$_x$N$_{1-x}$/diamond heterostructures (Cu/N/diamond and Cu/Cu-N/diamond) were chosen for building train and test datasets to ensure structural diversity. To collect training data, atomic vibrations were perturbed via both standard rattling and high-temperature MD rattling methods on these initial structures. Standard rattling was implemented by the `CompleteDFTvsMLBenchmarkWorkflow` function within `autoplex`[54,77,78] package (version 0.2.0). In parallel, high-temperature MD rattling was performed by `ASE`[25] within canonical (NVT) ensembles at 500 K. The train dataset was comprised of 32 snapshots of Cu, 34 snapshots of diamond, 140 snapshots of bare Cu/diamond interface, and 203 snapshots of the Cu/Cu$_x$N$_{1-x}$/diamond heterostructure ($x$ = 0 or 0.5). More information about these datasets can be found in **Supplementary Materials**.

Energies, atomic forces and stresses of these generated structures were obtained from DFT single-point calculations with `VASP`[75], managed by the `autoplex`. Specifically, DFT computations for rattled structures were performed via `DFTStaticLabelling` and `collect_dft_data` functions within `autoplex`. DFT labelling employed the projector augmented wave (PAW) method[79] and the PBEsol functional[80]. Moreover, we adopted Gaussian smearing of 0.05 eV width to electronic levels, a 600-eV cutoff for plane wave expansions, and a maximum spacing of 0.4 Å$^{-1}$ for meshing the reciprocal space. The total energy is attained with a convergence criterion of less than 10$^{-6}$ eV in the self-consistent electronic iterations.

**Table 1.** Comparison of the root-mean-square error (RMSE) for energies, forces and stresses of train dataset between the original `MACE-MPA-0` model and our finetuned `MACE-MPA-ft` model.

| MLIP models | $RMSE_{energy}$ (meV/atom) | $RMSE_{forces}$ (meV/Å) | $RMSE_{stress}$ (meV/Å$^3$) |
|---|---|---|---|
| `MACE-MPA-0` | 151.97 | 367.11 | 21.74 |
| `MACE-MPA-ft` | 0.63 | 17.60 | 1.20 |

During fine-tuning process, the MACE loss function $\mathcal{L}$—a weighted summation of Huber losses of energy, forces, and stress[60]—significantly decreased along with training iterations. Quality of the finetuned MACE potential was evaluated by comparing the supercell energies, atomic forces ($f_x, f_y, f_z$) and supercell stresses predicted by the MACE-MPA-ft models with those from DFT calculations. From **Fig. 2(a-c)**, it is obvious that MACE-MPA-ft accurately reproduces the supercell energies, atomic forces and supercell stresses from DFT calculations. For the train dataset, root-mean-square errors (RMSEs) of energy, forces and stresses are 0.63 meV/atom, 17.60 and meV/Å, 1.20 meV/Å$^3$, respectively. These low RMSE values confirm the quantum-accuracy of finetuned MACE models[25,81]. Meanwhile, we also calculated the errors of original pre-trained MACE-MPA-0 foundation model for the train dataset (in **Fig. 2(a-c)** and **Table 1**). Clearly, the precision of MACE-MPA-0 is insufficient for atomistic modelling and lattice dynamics simulation on Cu-diamond systems. The RMSEs of energies, forces and stress of our MACE-MPA-ft are only 0.4%, 4.8% and 5.5% of those of MACE-MPA-0 respectively, which verified the benefits of model fine-tunning[60].

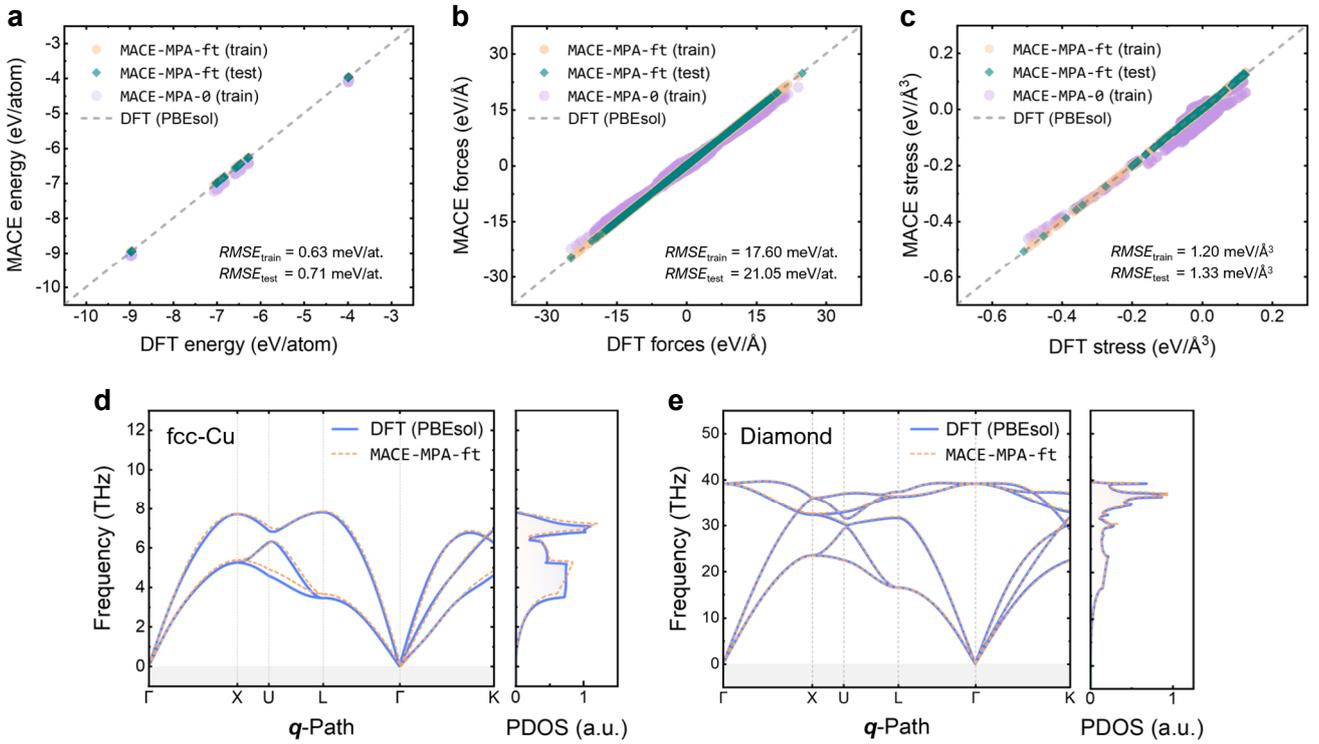

**Fig. 2.** Evaluation of the MACE models for Cu/diamond configurations involving nitrogen termination. Comparison of the **(a)** energies, **(b)** atomic forces and **(c)** stresses of train and test datasets predicted by the finetuned MACE model (MACE-MPA-ft), the pre-trained MACE model (MACE-MPA-0) and DFTs. Comparison of phonon dispersions and PDOS of **(d)** fcc-Cu and **(e)** diamond calculated by the MACE-MPA-ft and DFTs.

Effectiveness of the MACE-MPA-ft model for phononic calculations was validated by comparing to the DFT-predicted phonon dispersions and phonon density of states (PDOS), since accurately predicting phonon dispersions of a material is crucial for describing lattice dynamics. We calculated the second-order harmonic force constants (IFCs) through the finite displacement method based on ASE Calculator[25]. Combining the Phonopy package[67,68] with the second-order IFCs calculated from the MACE-MPA-ft and DFT, phonon dispersions of Cu and diamond were determined, as illustrated in **Fig. 2(d-e)**. Results show that MACE-MPA-

**11** / 27

`ft` potential accurately predicts the phonon frequencies at all high-symmetry points and accurately captures the dispersion behavior of each phonon branch. Meanwhile, PDOSs calculated by `MACE-MPA-ft` are almost identical to the DFT results as well.

### 3.2. Effects of N-termination on Cu/diamond interfacial thermal conduction

As discussed in **Introduction**, studies on enhancing the Cu/diamond ITC by surface modification mainly focus on metallic coatings, while the potential of non-metallic coatings has not been fully explored. Thereinto, nitrogen termination and functionalization of diamond have been applied to modulate the electronic and photonic performance, which should also be promising for tailoring the thermal transport at Cu/diamond interfaces. Hence, we focus on the tuning effect of N-termination of diamond for interfacial thermal conduction, to unveil the phonon-level mechanism of modulating heat transfer across Cu/diamond interfaces.

Consistent with the MACE fine-tuning datasets, two kinds of Cu/N$_x$/diamond heterostructures were built for lattice dynamics simulations: bare Cu/diamond interface ($x = 0$) and Cu/N/diamond interface ($x = 1$, full N-terminated). The bare Cu/diamond was constructed by directly stacking a 1×1×42-unit-cell Cu (fcc) supercell directly upon a 1×1×42-unit-cell diamond supercell along [001] direction, and Cu/N/diamond sandwich heterostructure was built by simply substituting the surface C-atoms in Cu/diamond heterostructure with N-atoms to mimic the structures of realistic N-terminated diamond[44,82]. Before performing lattice dynamics simulations, these Cu/N$_x$/diamond heterostructures were relaxed by `ASE` based on `MACE-MPA-ft` potential, until the maximal atomic force satisfied $f_{\max} < 0.02$ eV/Å. The relaxed structures are shown in **Fig. 3(a)**, with thicknesses of interfacial regions (gray areas) being ~5 Å.

ITCs of two kinds of interfaces were calculated through the RLD approach. A uniform sampling of phonon wave vectors in the first Brillouin zone on a 20×20×20 $q$-mesh grid was employed. To validate the robustness of RLD approach for C-N-Cu systems, we performed examination by interchanging the materials placed in the left and right "Lead" regions and deriving the corresponding ITC values. The resulting ITC variations were less than 4% attributed to numerical errors (**Fig. 3(b)**), thereby indicating no thermal rectification in such Cu/N$_x$/diamond systems. The ITC of bare Cu/diamond interface was determined as 41.6 MW/(m$^2$·K) by averaging the values derived from two directions (Cu ⇆ diamond), which was comparable with the NEMD-simulated ITCs for the flat Cu/diamond interfaces reported by Wang et al.[32]. The ITC of Cu/N/diamond interface was calculated to be 50.3 MW/(m$^2$·K) by the same way, exhibiting an ~21% enhancement enabled by the N-termination on diamond surface. Furthermore, we also conducted the convergence tests against in-plane supercell sizes and $q$-mesh grids of the first Brillouin zone (see **Fig. A1** in **Appendix A**), which corroborated that the in-plane size of 1×1 and *q*-mesh of 20×20×20 were both sufficient for the Cu/N$_x$/diamond systems under investigation[64]. The results reveal that nitrogen-termination engineering can function as an effective strategy for enhancing the Cu/diamond interfacial thermal conduction, which opens potential venues for tuning thermal properties of Cu/diamond composites.

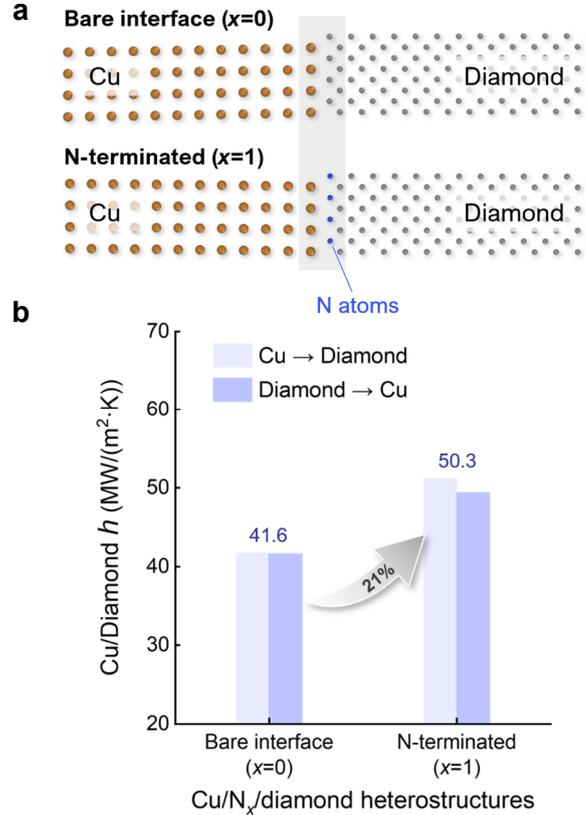

**Fig. 3.** The Cu/N$_x$/diamond heterostructures and the corresponding interfacial thermal conductance. **(a)** Relaxed structures of two kinds of Cu/N$_x$/diamond interfaces with different N-termination, i.e., bare Cu/diamond ($x = 0$) and Cu/N/diamond ($x = 1$, full N-terminated interlayer), respectively. **(b)** RLD calculations reveal the tuning effects of ITC enabled by varying the N-termination, i.e., a 21% improvement in Cu/diamond ITC for $x = 1$.

Cumulative ITCs of the two Cu/N$_x$/diamond interfaces were determined (**Fig. 4(a)**). Cu/N/diamond interface exhibits better thermal conductance than bare Cu/diamond for phonon frequencies above 4 THz. Since the only structural difference between the two heterostructures is the N-terminated interlayer, both the phonon population and the energy flux can be considered identical for bare Cu/diamond and Cu/N/diamond interfaces under the harmonic approximation. Therefore, the decisive factor governing the difference in ITC is the spectral phonon transmittance $\mathcal{T}(\omega)$. As shown in **Fig. 4(b)**, $\mathcal{T}(\omega)$ of Cu/N/diamond interface is higher than bare Cu/diamond for phonon frequencies above 4 THz, while the transmittance below 4 THz is almost identical for the two interfaces.

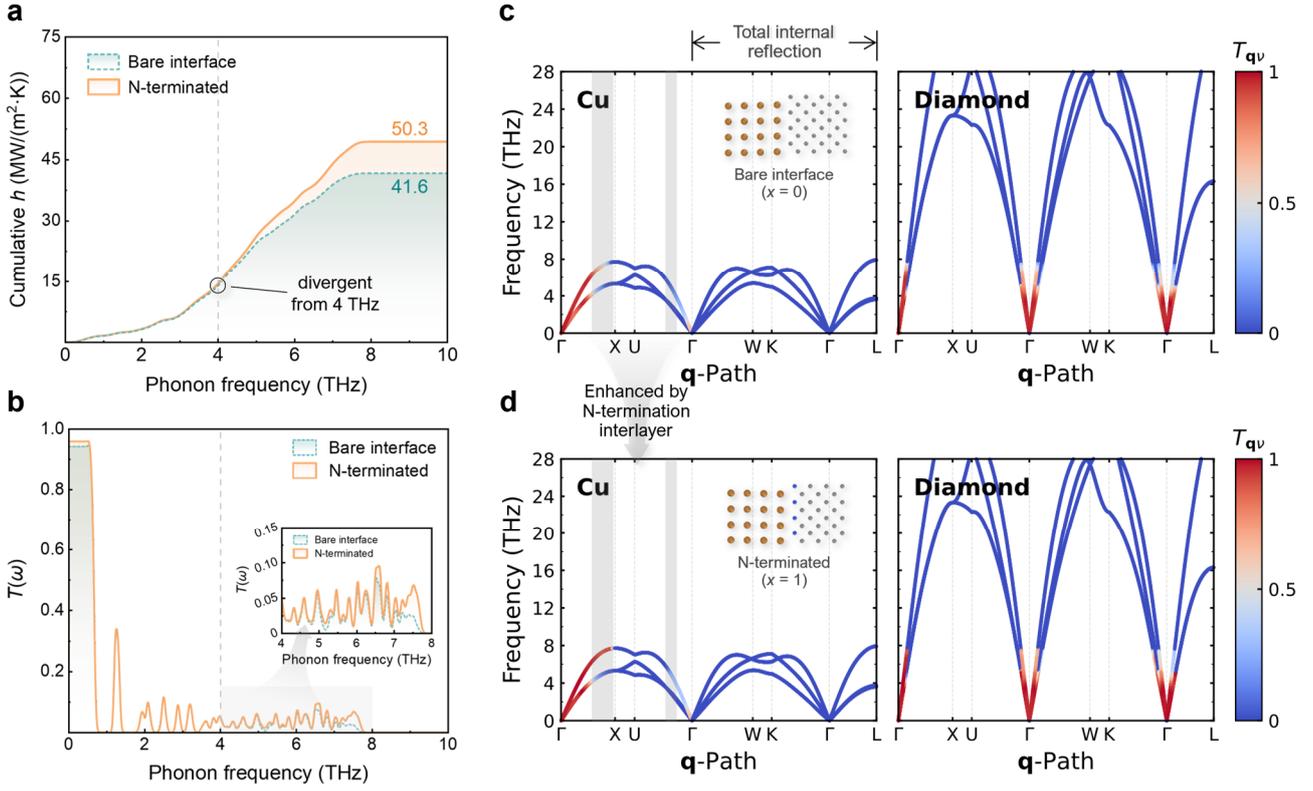

**Fig. 4.** Interfacial thermal conduction and phonon transmission across the two kinds of Cu/N$_x$/diamond interfaces. **(a)** Cumulative ITCs of Cu/N$_x$/diamond heterostructures with two different compositions of nitrogen-termination ($x = 0, 1$), and the ITC of Cu/N/diamond heterostructure ($x = 1$, full N-terminated) is better than bare Cu/diamond ($x = 0$) above 4 THz, that agrees with **(b)** Spectral phonon transmittance across the two interfaces. Mode-resolved phonon transmittance mapping across **(c)** the bare Cu/diamond interface ($x = 0$) and **(d)** the Cu/N/diamond interface ($x = 1$). LA phonons of fcc-Cu along $\varGamma$–$X$, $\varGamma$–$U$ paths are directly tuned by introducing diamond's N-terminations. Transmittances of phonons in fcc-Cu along other high-symmetry paths with larger incident angles are invariably zero, due to the total internal reflection.

Building upon these observations, we further analysed the mode-resolved phonon transmittance to elucidate phonon transport behaviors across interfaces. The first Brillouin zones in reciprocal space of fcc-Cu and diamond are both truncated octahedra, and we selected the typical high-symmetry path $\varGamma$–$X$–$U$–$\varGamma$–$W$–$K$–$\varGamma$–$L$ around $z$-axis for analysis[83]. Due to the anisotropy of phonon wavevectors, total internal reflection may occur when phonons propagate from a low-group-velocity medium to a high-group-velocity one with exceeding the critical angle $\theta_c$. Mode-resolved phonon transmittances of bare Cu/diamond and Cu/N/diamond interfaces are illustrated in **Fig. 4(c, d)**. Note that phonon dispersions of the diamond above 28 THz are omitted for clarity, as the transmittance of these high-frequency phonons is trivially zero within the harmonic approximation employed by the RLD method.

It is clear that the phonon transmittances of Cu along $\varGamma$–$L$, $\varGamma$–$K$, and $\varGamma$–$W$ paths vanish, and the N-termination on diamond does not alter this behavior. This can be attributed to the larger incident angles along these paths. Since the phonon group velocities in Cu are significantly lower than those in diamond, the critical angle $\theta_c$ is small, causing phonons incident at larger angles to undergo total internal reflection. For the Cu/N/diamond interface, the N-interlayer enhances transmittance for phonons incident at smaller angles, e.g., along the $\Gamma$–$X$, $\Gamma$–$U$ paths. In particular, the longitudinal acoustic (LA) phonons transmittance above 4 THz—especially along the $\Gamma$–$X$, $\Gamma$–$U$ paths—are significantly increased after introducing the N-termination on diamond surface.

### 3.3. Tuning phonon transport by interfacial nitrogen

The ITC of heterostructure is strongly correlated with the phonon spectra of the two materials, especially the matching of local vibrational states near the interface[38,84], which can be characterized by the LDOS overlap of the materials on both sides of interface[9]. As illustrated in **Fig. 5(a)**, we concentrated on the LDOSs of the first layer of primitive cell in Cu and diamond (calculated by **Eq. (7)**) that are most responsible for the phonon transport across interface[85]. The first layer of primitive cell in Cu and diamond consists of one Cu atom and two C atoms for the bare Cu/diamond interface, or one Cu, N, C atom for Cu/N/diamond interface respectively. Note that the LDOS of 1st diamond layer was summed by the two C atoms ($C_{(1)}$, $C_{(2)}$ shown in **Fig. 5(a)**) or N and C atoms. In **Fig. 5(b, c)**, we found that replacing the superficial $C_{(1)}$ atoms with N caused a blueshift of the optical phonon branches of diamond, that the LDOS peaks reduced from 33–37 THz to 28–31 THz. This corresponds to the atomic mass modification of the diamond surface.

To quantify the mass-modification effects on enhancing the LDOS overlap by N-termination, we referred to the DOS overlapping factor $S$ discussed in literature,[86–88]

$$S = \frac{\int_0^{+\infty} D_1(\omega) D_2(\omega) d\omega}{\left[\int_0^{+\infty} D_1(\omega) d\omega\right]\left[\int_0^{+\infty} D_2(\omega) d\omega\right]}, \quad (11)$$

where $D_1$ and $D_2$ are the LDOS of atoms on both sides of the interface. Values of the overlapping factor $S$ are related to the phonon coupling between two layers. As a result, the $S$ value increases 13.7% from 0.0168 for bare Cu/diamond interface to 0.0191 for Cu/N/diamond interface. Thus, the LDOS overlap condition between the first layer of Cu and diamond is improved by introducing N-termination, while more channels are available for heat transport between the two layers.

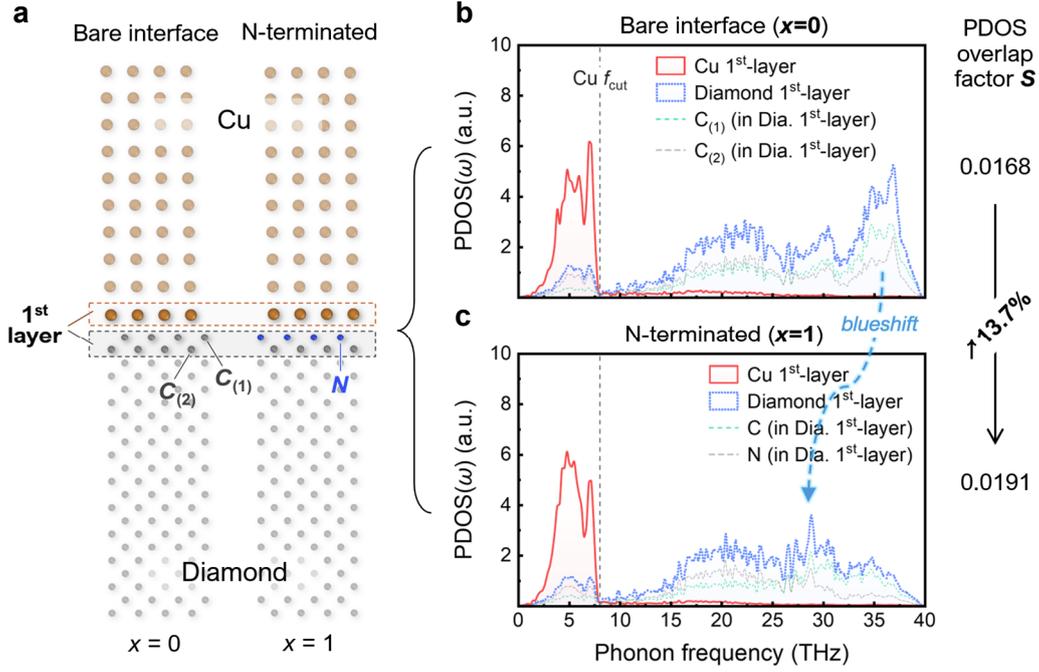

**Fig. 5.** Interfacial vibrational spectra of two kinds of Cu/N$_x$/diamond interfaces and the corresponding LDOS analysis. **(a)** The first layer of primitive cell in Cu or diamond is selected for calculating LDOS. LDOS spectra of **(b)** bare Cu/diamond interface ($x = 0$) and **(c)** Cu/N/diamond interface ($x = 1$, full N-terminated). Replacing the superficial C$_{(1)}$ atom with N caused a blueshift of the optical phonons of diamond surface. The overlapping factor increases 13.7% from bare interface to N-terminated interface, showing better phonon coupling between Cu and diamond layers.

Besides the mass-modification effects, N-termination engineering is effective in tuning the surface energy states[44,82] or adsorption energies[47], which should be promising to affect the interfacial bonding conditions thereby tuning heat transfer between Cu and diamond layers[70]. We here conducted COHP analysis to unravel the bond tuning effects induced by N-termination. All the DFT-SCF calculations were performed on 30-Å slabs sliced from the interfacial region of Cu/N$_x$/diamond heterostructures to save the computational costs. After the `VASP` iterations were converged, the `LOBSTER` post-processing was then implemented on the first layer of primitive cell in Cu and diamond as the foregoing LDOS analysis did, where the chemical bonds are deemed responsible for the interfacial thermal conduction as well. We only concerned the bonding between atoms with different $z$-coordinates, since $x$, $y$-directions were periodic boundaries and interfacial thermal conduction was along the $z$-direction.

The COHP spectra and ICOHP values were derived for both the bare Cu/diamond and Cu/N/diamond interfaces, which were averaged by the number of all bonds within the first Cu and diamond layers (**Fig. 6(a)**). The N-terminated interface shows a 13.4% stronger bonding indicated by the ICOHP, attributing to the significant stronger interactions between -16–-23 eV shown in the COHP spectrum. We further calculated the COHP spectra and ICOHP values of C$_{(1)}$-Cu, N-Cu, C$_{(1)}$-C$_{(2)}$ and C-N interactions respectively (in **Fig. 6(b, c)**).

The difference between $C_{(1)}$-Cu and N-Cu interactions below $\varepsilon_F$ is negligible comparing to $C_{(1)}$-$C_{(2)}$ and C-N interactions. It is obvious that the C-N interactions contribute most to the deviation in -16–-23 eV for two interfaces, implying that the stronger C-N interactions are responsible for enhancing ITC. In summary, the N-termination engineering proves to be effective in tuning the interfacial heat conduction via the surficial mass modification and bonding regulation effects.

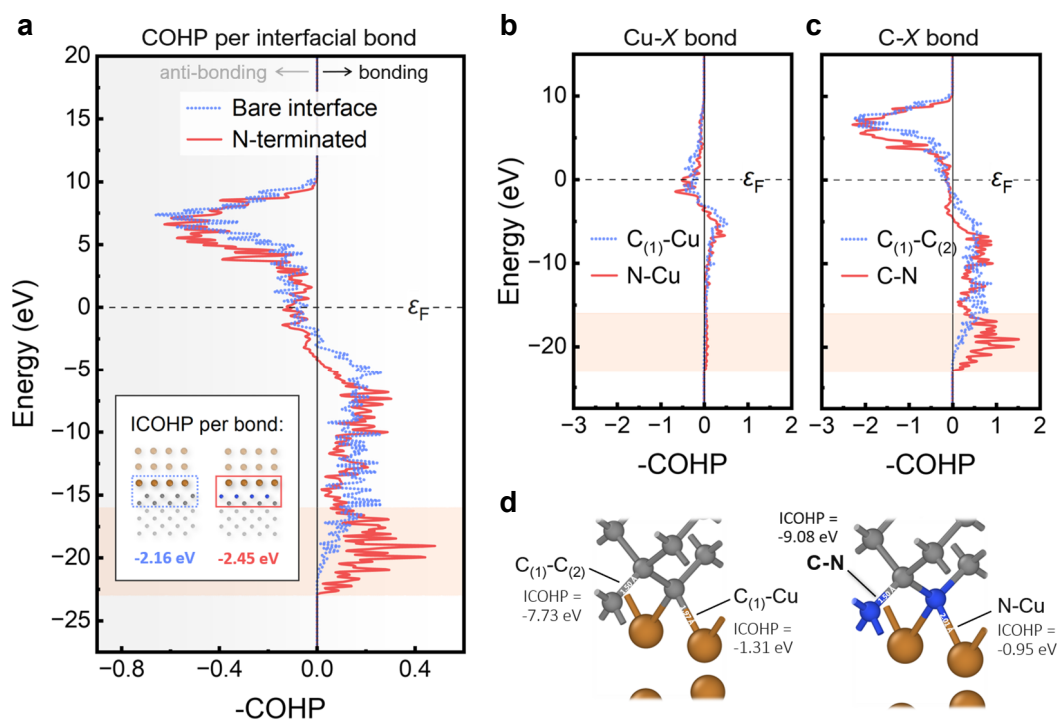

**Fig. 6.** Interfacial bonding conditions and COHP analysis of the two kinds of Cu/N$_x$/diamond interfaces. **(a)** COHP spectra of the bare Cu/diamond and Cu/N/diamond interfaces ($\varepsilon_F$: Fermi energy), and the inset plot gives their ICOHP values. Note that the COHP and ICOHP are both statistical values of the first layer of primitive cell in Cu and diamond (blue and red rectangle shown in the inset). The Cu/N/diamond (N-terminated) interface exhibits a 13.4% better bonding strength, contributing to the significant stronger interactions between -16–-23 eV. The COHP spectra of **(b)** $C_{(1)}$-Cu, N-Cu interactions and **(c)** $C_{(1)}$-$C_{(2)}$, C-N interactions, and the C-N interactions contribute most to the deviation between -16–-23 eV for two interfaces. **(d)** The ICOHP values and bond lengths of the four different bond-pairs, showing the stronger C-N interactions are responsible for enhancing ITC.

# 4. Conclusion

In this work, we investigated tuning effects of surface termination engineering of diamond on Cu/diamond interfacial thermal transport. Focusing on the nitrogen-terminated system, we finetuned the pre-trained `MACE-MPA-0` foundation model to the quantum-accurate `MACE-MPA-ft` model using customized C-N-Cu training datasets. Lattice dynamics simulations based on `MACE-MPA-ft` revealed that the N-termination of diamond enhances the Cu/diamond interfacial thermal conductance by approximately 21%. Mode-resolved phonon spectroscopy showed a significant enhancement in the transmittance of LA phonons above 4 THz—particularly along the $\Gamma$–$X$ and $\Gamma$–$U$ paths. Furthermore, we conducted local vibrational states and interfacial bonding analyses, characterized by the LDOS overlap and COHP spectra, which demonstrated that the N-termination engineering effectively tunes the interfacial heat conduction through surface mass modification and bonding regulation effects. These findings offer fundamental insights into the role of nitrogen in tuning thermal transport at Cu/diamond interfaces and provide a feasible strategy for designing high-performance Cu-diamond composites via surface nitrogen-termination.

# Appendix A. Convergence tests of lattice dynamics

We conducted convergence tests of lattice dynamics simulations, against *q*-mesh grids of the first Brillouin zone and in-plane supercell sizes (**Fig. A1**). It revealed that the *q*-mesh of 20×20×20 and in-plane size of 1×1 used in our model settings were both sufficient for calculating the Cu/diamond ITCs and eliminating the unphysical thermal rectification phenomenon, agreeing with the reported studies[64].

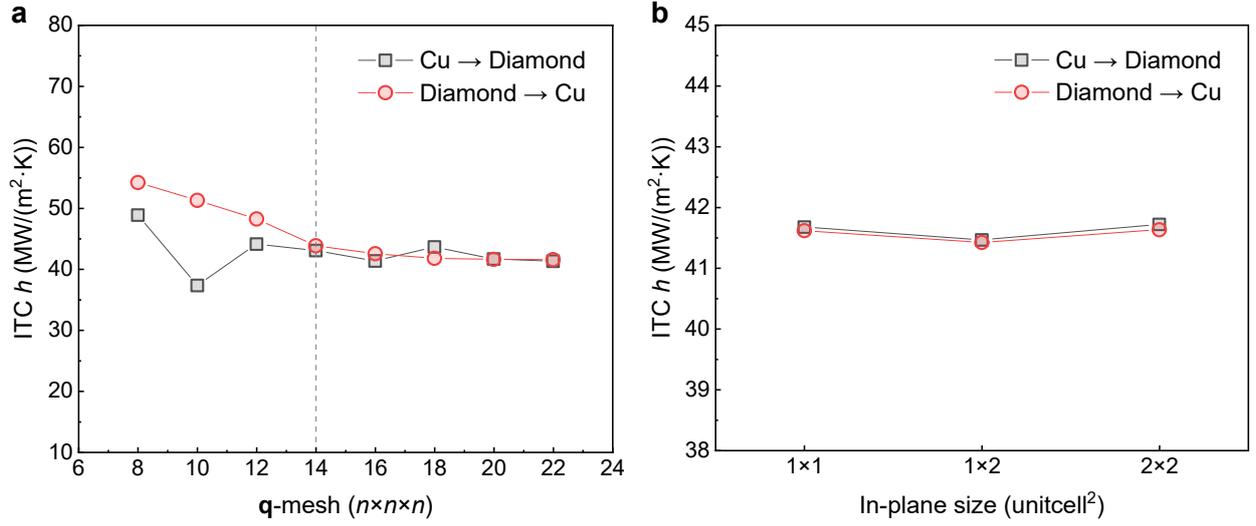

**Fig. A1.** Convergence tests of lattice dynamics against (a) *q*-mesh of the 1st Brillouin zone, (b) in-plane supercell sizes.

## Acknowledgements

This work was financially supported by the National Science and Technology Major Project (Grant No. 2025ZD0803800).

## Competing interest

The authors declare that they have no known competing financial interests or personal relationships that could have appeared to influence the work reported in this paper.

## Data availability

The `MACE-MPA-0` foundation model is available on https://github.com/ACEsuit/mace-foundations. Train and test datasets for MACE fine-tuning, and files of the finetuned `MACE-MPA-ft` model are available on https://github.com/YangLight/Supplementary_Materials_-JCP26-AR-00623-. Data supports the findings of this study are available from the corresponding author upon reasonable request.